\documentclass[]{aastex631}

\shorttitle{Radio Counterpart of V2187 Cyg}
\shortauthors{Rodr\'\i guez et al.}

\graphicspath{{./}{figures/}}%\usepackage{paralist}
\defcitealias{2023A&A...678A.185R}{RLCG23}

\defcitealias{1983RMxAA...8..163R}{RC83}

\newcommand{\be}{\begin{equation}}
\newcommand{\ee}{\end{equation}}

\shorttitle{Radio Emission from V2187 Cyg}
\shortauthors{Rodr\'iguez, Lizano, Cant\'o, Gonz\'alez \& Tapia}

\begin{document}

\title{Understanding the Radio Emission from the $\beta$ Cep star V2187 Cyg}

\author[0000-0003-2737-5681]{Luis F. Rodr\'{\i}guez}
\affiliation{Instituto de Radioastronom\'{\i}a y Astrof\'{\i}sica\\
Universidad Nacional Aut\'onoma de M\'exico, Apdo. Postal 3-72, Morelia, Michoac\'an 58089, Mexico}
\affiliation{Mesoamerican Centre for Theoretical Physics\\
Universidad Aut\'onoma de Chiapas, Tuxtla Guti\'errez, Chiapas 29050, Mexico}
\author[0000-0002-2260-7677]{Susana Lizano}
\affiliation{Instituto de Radioastronom\'{\i}a y Astrof\'{\i}sica\\
Universidad Nacional Aut\'onoma de M\'exico, Apdo. Postal 3-72, Morelia, Michoac\'an 58089, Mexico}
\author[0000-0003-3863-7114]{Jorge Cant\'o}
\affiliation{Instituto de Astronom\'{\i}a\\
Universidad Nacional Aut\'onoma de M\'exico, Apdo. Postal 70-264, CDMX 04510, Mexico}
\author[0009-0000-0711-7111]{Ricardo  F. Gonz\'alez}
\affiliation{Instituto de Radioastronom\'{\i}a y Astrof\'{\i}sica\\
Universidad Nacional Aut\'onoma de M\'exico, Apdo. Postal 3-72, Morelia, Michoac\'an 58089, Mexico}
\author[0000-0002-0506-9854]{Mauricio Tapia}
\affiliation{Instituto de Astronom\'{\i}a\\ Universidad Nacional Aut\'onoma de M\'exico, Ensenada, B. C., CP 22830, Mexico}

\begin{abstract}

We analyze the radio emission from the $\beta$ Cep star V2187 Cyg using archive data from the Jansky Very Large Array.
The observations were made in ten epochs at 1.39 and 4.96 GHz in the highest angular resolution A configuration.
We determine a spectral index of of $\alpha = 0.6\pm0.2$ ($S_{\nu} \propto \nu^\alpha$), consistent with an ionized wind 
 or a partially optically-thick synchrotron or gyrosynchrotron source. The emission is spatially unresolved at both frequencies.
The 4.96 GHz data shows a radio pulse with a duration of about one month that can be modeled in terms of an 
internal shock in the stellar wind produced by a sudden increase in the mass-loss rate and the terminal velocity.
The quiescent radio emission of V2187 Cyg at 4.96 GHz (with a flux density of $\simeq 150~\mu Jy$), cannot be explained in terms 
of an internally (by V2187 Cyg) or externally (by a nearby O star) photoionized wind.
We conclude that, despite the spectral index suggestive of free-free emission from an ionized wind,  the radio emission of V2187 Cyg most likely has a magnetic origin, a possibility that can be tested with
a sensitive search for circular polarization in the radio,
as expected from gyro-synchrotron radiation, and also by trying to measure the stellar magnetic field, that
is expected to be in the range of several kGauss.

\end{abstract}



%\keywords{Radio continuum emission(1340) --- K dwarf stars(879)}

\section{Introduction} \label{sec:intro}

Main sequence stars in the B2-B3 spectral type range are not expected to be bright thermal (free-free) radio sources
given their small ionizing photon rate ($10^{43.9-44.9}$ s$^{-1}$; \cite{1973AJ.....78..929P}). 
In the most favorable case of optically-thin free-free emission, these rates will produce sources with centimeter flux densities
in the range of $\simeq$0.9--9.0 mJy at a distance of 1 kpc (\cite{1983RMxAA...8..163R}; hereinafter RC83). However, these stars have winds that 
will absorb the ionizing photons and produce much fainter (by orders of magnitude), partially optically-thick 
free-free sources. In the case that the ionizing photon rate
matches the value needed to fully ionize the wind, these sources will
have weak centimeter flux densities
in the range of $\simeq$0.01--1 mJy at a distance of 1 kpc  (\citetalias{1983RMxAA...8..163R} ).  

V2187 Cyg is a $\beta$ Cep star located at a distance of 1.78 kpc (DR3 Gaia parallax $\mu = 0.561 \pm 0.016$ mas; 
Gaia collaboration et al. 2016; 2023) with a B2--B3 spectral type
and time variable centimeter radio emission in the range of
$\simeq$0.2--1.2 mJy (Tapia et al. 2014). These flux densities are high above those expected for an ionized wind. 
In this paper we present an analysis of archive Very Large Array data of this star and discuss alternatives
that could explain its relatively large radio continuum flux density.

\section{Observations} 
\label{sec:observations}

We have used the observations of the project 10C-134 from the archives of the 
Karl G. Jansky VLA of NRAO\footnote{The National 
Radio Astronomy Observatory is a facility of the National Science Foundation operated
under cooperative agreement by Associated Universities, Inc.}. These observations were made at 1.39 and 4.96 GHz
with a total bandwidth of 256 MHz in 10 epochs between 2011 March 27 and 2011 August 28.
The data were acquired in the A or B configurations pointing toward the star Cyg OB2 \#9.
Even when V2187 Cyg is located at $\sim 2\rlap.'9$ from Cyg OB2 \#9
we could obtain images of good quality over the full extent of the
primary beams ($\sim 30'$ at 1.39 GHz and $\sim 10'$ at 4.96 GHz). 
This was possible because the Jansky VLA acquires the continuum data in
narrow channels (2 MHz), that do not produce significant bandwidth smearing. For all observations J2007+4029
was the gain calibrator and J0542+498 (3C147) the amplitude calibrator. 
The data were calibrated in the standard manner using the CASA (Common Astronomy Software Applications;  McMullin et al. 2007) package of NRAO and
the pipeline provided for VLA\footnote{https://science.nrao.edu/facilities/vla/data-processing/pipeline} observations. We made images using a robust weighting (Briggs 1995) of 0 to
optimize the compromise between angular resolution and sensitivity.  All images were also corrected for the primary beam
response. 

\begin{deluxetable*}{ccc}
\tablenum{1}
\tablecaption{4.96 GHz observations of V2187 Cyg from project 10C-134}
\tablewidth{900pt}
\tabletypesize{\scriptsize}
\tablehead{
\colhead{Epoch}  & \colhead{VLA}  & \colhead{Flux Density}  \\
\colhead{(MJD)} & \colhead{Configuration} & \colhead{($\mu$Jy)}  }
\decimalcolnumbers
       \startdata
55647.78 &  B &  150$\pm$18 \\
55654.77 &  B &  329$\pm$19 \\
55683.66 &  B &  236$\pm$18 \\
55701.59  &  B &  143$\pm$12 \\
55728.62  &  A &  176$\pm$15 \\
55739.57  &  A &  205$\pm$20 \\
55749.56 &  A &  157$\pm$16 \\
55769.44 &  A &  204$\pm$17 \\
55785.36 &  A &  157$\pm$12 \\
55801.31  &  A &  194$\pm$20 \\
\enddata  
\label{tab:param1}
\end{deluxetable*}

%Finally, the images were also corrected for wide-field effects
%using the gridding option  \it widefield \rm with 10$\times$10 subregions in the task TCLEAN. 
 
\begin{figure*}[!t]
%  \hskip-1.5cm
\vskip-1.0cm
\hskip-1.0cm\includegraphics[width=0.52\linewidth]{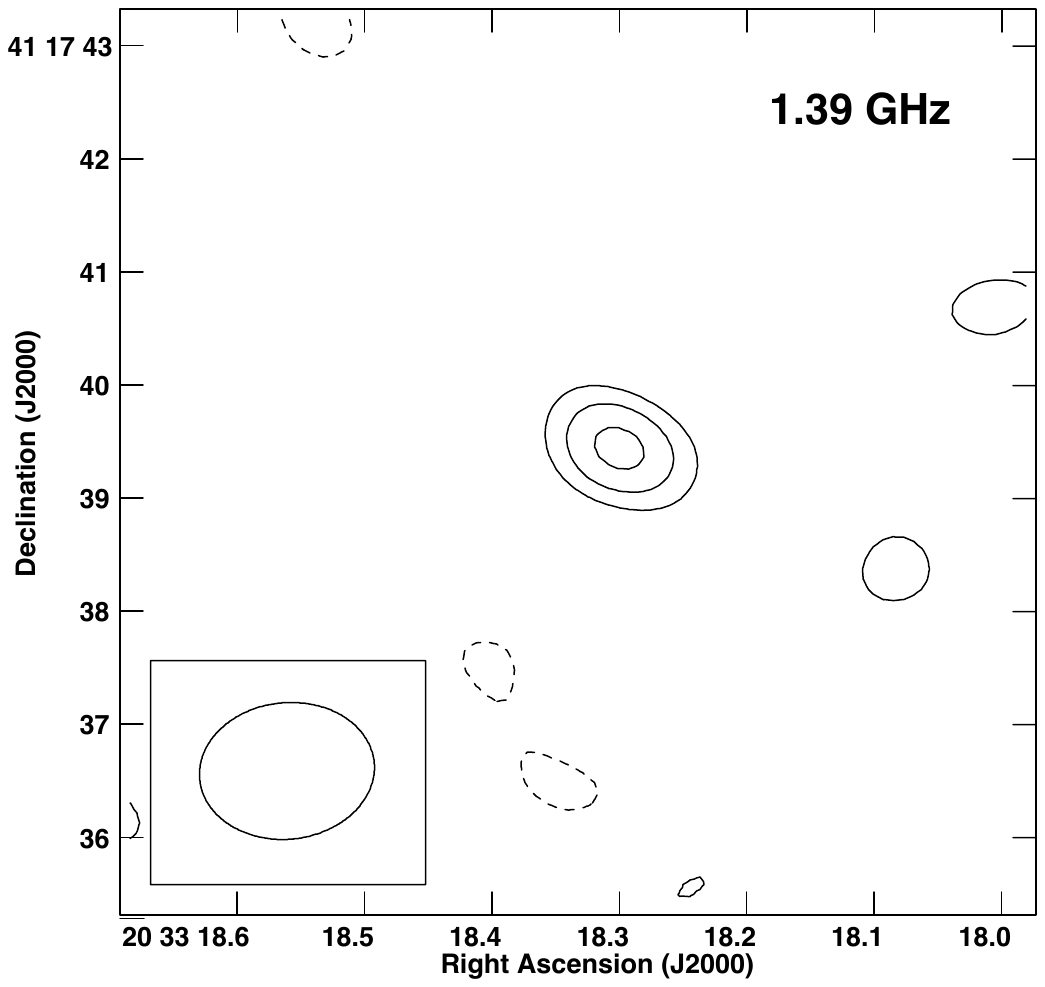}\hskip-1.0cm\includegraphics[width=0.52\linewidth]{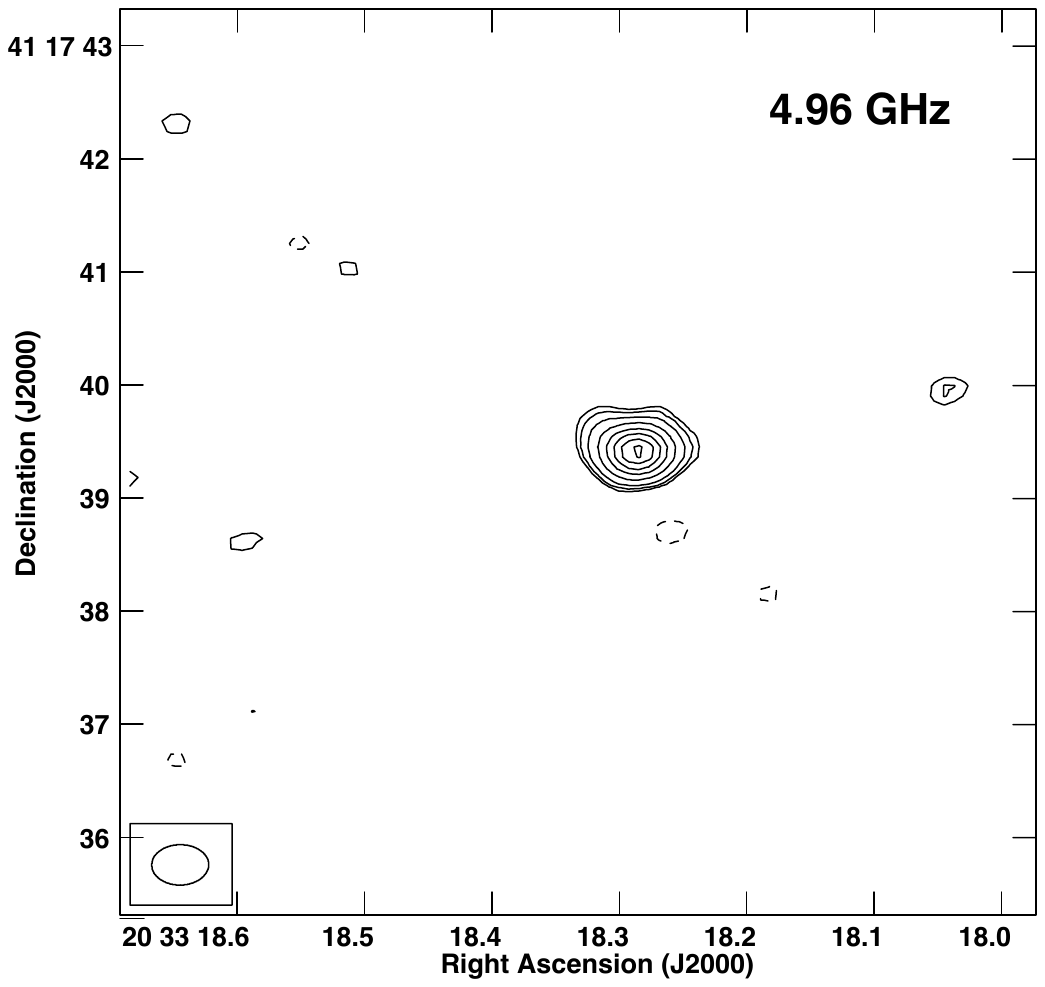}%\vskip-1.8cm
\vskip-1.8cm
\caption{Very Large Array contour images of V2187 Cyg
at 1.39 (left) and 4.96 GHz (right). For the 1.39 GHz image the contours 
are -3, 3, 4 and 5
times 11.5  $\mu$Jy beam$^{-1}$, the rms noise in this region of the image. The synthesized beam 
($1 \rlap.{"}55 \times 1\rlap.{"}21; -84\rlap.^\circ2$)
 is shown in the bottom left corner of the respective image. For the 4.96 GHz image the contours 
 are -3, 3, 4, 6, 10, 15, 20, 25 and 30
 times 5.6  $\mu$Jy beam$^{-1}$, the rms noise in this region of the image. The synthesized beam 
 ($0\rlap.{"}50 \times 0\rlap.{"}36; -89\rlap.^\circ3$)
 is shown in the bottom left corner of the respective image.
 }
\label{fig:widefig2}
\end{figure*}

 In Figure \ref{fig:widefig2} we show contour images of V2187 Cyg at the two frequencies observed, made from concatenating all
 epochs. The flux densities 
 measured for V2187 Cyg were 104$\pm$31 and 210$\pm$13 $\mu$Jy at   1.39 and 4.96 GHz, respectively. 
 The quoted errors are largely due to variability, and not measurement uncertainties. These values
 imply an average spectral index of $\alpha = 0.6\pm0.2$ ($S_{\nu} \propto \nu^\alpha$), consistent with an ionized wind 
 (\cite{1975A&A....39....1P} ; \cite{1986ApJ...304..713R} ) 
 or a partially optically-thick synchrotron or gyrosynchrotron source (\cite{1985ARA&A..23..169D}; \cite{2021MNRAS.507.1979L}).
 
%\noindent{\bf \color{red} Table 1 here}

\begin{deluxetable*}{ccccccccc}
\tablenum{2}
\tablecaption{Adopted Parameters for Main Sequence Early-Type B Stars}
\tablewidth{900pt}
\tabletypesize{\scriptsize}
\tablehead{
 \colhead{}    & \colhead{T$_\star$} & \colhead{L$_\star$} & \colhead{R$_\star$} &  \colhead{$\dot{\rm N}_i$}
   & \colhead{$\dot{\rm N}_{wind}$} 
       & \colhead{M$_\star$} & \colhead{$\dot {\rm M}$} & \colhead{${\rm v}_\infty$} \\ 
\colhead{Sp} &  \colhead{(K)} &  \colhead{(L$_\odot$)} 
 & \colhead{(R$_\odot$)} & \colhead{(s$^{-1}$)}
      & \colhead{(s$^{-1}$)} &\colhead{(M$_\odot$)} & \colhead{(M yr$^{-1}$)} & \colhead{(km s$^{-1}$)}
      }
\decimalcolnumbers
       \startdata
      B0V & 30900 & 4.8$\times10^4$ &  
      7.6  & 4.0$\times$10$^{47}$ & 1.7$\times$10$^{42}$ & 16.6  &  5.2$\times$10$^{-8}$ & 2440 \\ 
     B0.5V & 26200 & 2.0$\times10^4$ & 7.1 &
        3.2$\times$10$^{46}$ & 2.2$\times$10$^{41}$   & 12.3 & 1.5$\times10^{-8}$  & 2100 \\
       B1V & 22600 &  8.9$\times10^3$ & 6.2 &
        3.2$\times$10$^{45}$ &  2.8$\times$10$^{40}$  & 9.2 & 4.9$\times10^{-9}$ & 1710 \\
       B2V & 20000 & 5.0$\times10^3$ & 5.6 &
       6.3$\times$10$^{44}$ & 2.6$\times$10$^{40}$  & 7.4 & 2.2$\times10^{-9}$ & 1340 \\
     B3V &  17900 &  1.7$\times10^3$ & 4.4 &
     7.9$\times$10$^{43}$ & 1.5$\times$10$^{39}$ & 5.1 & 4.3$\times10^{-10}$ & 1090 \\ 
     \enddata  
\tablecomments{The references for the parameters are discussed in Section 3.4. Sp = Spectral Type, T$_\star$ = Temperature, L$_\star$ = Luminosity,
       R$_\star$ = Radius, $\dot {\rm N_i}$ = Ionizing Photon
Rate, $\dot {\rm N}_{wind}$ = Ionizing Photon Rate required to fully ionize the wind, 
M$_\star$ = Mass, $\dot {\rm M}$ = Wind Mass Loss Rate, v$_\infty$ =  Wind Terminal Velocity.}
\label{tab:param}
\end{deluxetable*}

\section{Discussion}
\label{sec:discussion}

\subsection{The angular size of the radio emission}

 We used all ten observations made in the 10C-134 project as listed in Table 1. This Table shows the epoch (in Modified Julian Dates), the VLA configuration and the flux density at 4.96 GHz.
The first four observations were made in the B configuration and the last six in the A configuration. These last six observations seem to be tracing the quiescent state of the radio emission from V2187 Cyg, while the observations of the first four epochs relate to a  radio pulse. The observations at 1.39 GHz are noisy and we could not analyze the individual epochs, using only the
concatenation of all 10 epochs.

To estimate the angular size of the radio emission we used the data from the second epoch (taken to trace the pulse state)
and the concatenation of the last six epochs (taken to trace the quiescent state). We subtracted the additional sources
in the field by cleaning them with the CASA task TCLEAN and removing their clean components with the task UVSUB.
We then centered the (u,v) data at the position of V2187 Cyg with the task FIXVIS. Finally, the (u,v) data was fitted with a
circular Gaussian function using UVMODELFIT. For the second epoch we obtain 
a 3-$\sigma$ upper limit of $\leq 0\rlap.''36$. For the concatenation of the
last 6 epochs we obtain a 3-$\sigma$ upper limit of $\leq 0\rlap.''14$ for the angular size of V2187 Cyg 
(249 au at a distance of 1.78 kpc).

\subsection{Comparison with the Gaia position}

To corroborate the coincidence of the radio emission with the star, we used the highly accurate Gaia position for
V2187 Cyg (Gaia collaboration et al. 2016; 2023). The average epoch of the radio observations is 2011.551. Correcting the Gaia
position for its proper motions we obtain a position of $RA(J2000) = 20^h~ 33^m 18\rlap.^s280; 
DEC(J2000) = 41^\circ~ 17'~ 39\rlap.{''}39$
for that epoch. The 4.96 GHz position for the same epoch is $RA(J2000) = 20^h~ 33^m 18\rlap.^s286 \pm 0\rlap.^s002; 
DEC(J2000) = 41^\circ~ 17'~ 39\rlap.{''}42 \pm 0\rlap.{''}01$, confirming the close association of the radio and optical emissions,
since their positions coincide within $0\rlap.{''}03$.

\subsection{Periodic variability in the radio emission?}

In Tapia et al. (2014) evidence was presented in favor of a periodic variability in the radio emission from V2187 Cyg.
The period of the radio emission was 12.8 days and it was detected in observations taken from 1984 to 2005. However,
when we add the 2010 data the statistical significance of the periodicity disappears. We also failed to find periodicities in the 2010 data.

\subsection{Interpretation of the 4.96 GHz pulse}

In this section, we explore the possibility that the observed radio pulse 
at 4.9 GHz could be due to shocks
in the stellar wind of V2187 Cyg.  \cite{2023A&A...678A.185R} (hereafter RLCG23) modeled a
similar radio pulse observed at 10 GHz in the star $\epsilon$ Eridani as produced
by internal shocks in the stellar wind. The model assumes that  the stellar wind ejection velocity  suddenly increases producing 
a two-shock structure, called the working surface (WS), by the interaction of a fast perturbed wind that rams  into
the steady wind of the star.  The wind mass-loss rate can also change during this event. 
The WS then emits the energy dissipated in this process. Here we assume that  V2187 Cyg suffered a similar violent event.

To model the radio pulse, \citetalias{2023A&A...678A.185R} 
followed the work of \cite{2000MNRAS.313..656C} and   \cite{2022MNRAS.509.1892M}
for the simple case of step function variations of the wind velocity and mass-loss rate. In this case, the WS  undergoes 
an initial phase with constant velocity $v_{WS0}$, when it is bounded by two shock fronts,  followed by a phase of deceleration with $v_{ws}(t)$, when the shock front closest to the star disappears. 
Equations (1) and (2) of RLCG23  describe the evolution of the pulse bolometric luminosity: first, during the constant velocity phase, the 
luminosity is constant and is given by 
\begin{equation}
 L_{WS} = \left({\Omega \over 4 \pi}\right) {{1}\over{2}} \dot{M}\,v_w^2\,
    \biggl[{{b}\over{a}}\biggl(a-{{v_{WS0}}\over{v_w}}\biggr)^3 +
    \biggl({{v_{WS0}}\over{v_w}}-1\biggr)^3\biggr]\,, \quad t \le t_c,
    \label{eq:lumi}
\end{equation}
and later, during the deceleration stage, the luminosity decreases as
\begin{equation}
L_{WS}(t) = \left({\Omega \over 4 \pi}\right) {{1}\over{2}} \dot{M}\,v_w^2\, 
    \biggl[{{v_{WS}(t)}\over{v_w}}-1\biggr]^3\,, t > t_c,
	\label{eq:lumf}
\end{equation}
where $\Omega$ is the solid angle of stellar wind where the velocity variation occurs, and
 $a$ and $b$ are, respectively, the wind velocity and mass-loss rate variation factors that last only for an ejection time $\Delta \tau$. The initial constant WS velocity is $v_{WS0} = \sigma v_w$, where $\sigma = a^{1/2} ( 1+a^{1/2} b^{1/2})/(a^{1/2}+b^{1/2})$, and the  WS velocity during the deceleration phase, $v_{ws}(t)$, is given by equation (2) of \cite{2022MNRAS.509.1892M}. The constant velocity phase will last a time $t_c$ given by $t_c = a  \Delta \tau/(a-\sigma)$. % \Delta \tau $.

\begin{figure}
\vskip-3.0cm
\includegraphics[width=0.8\columnwidth]{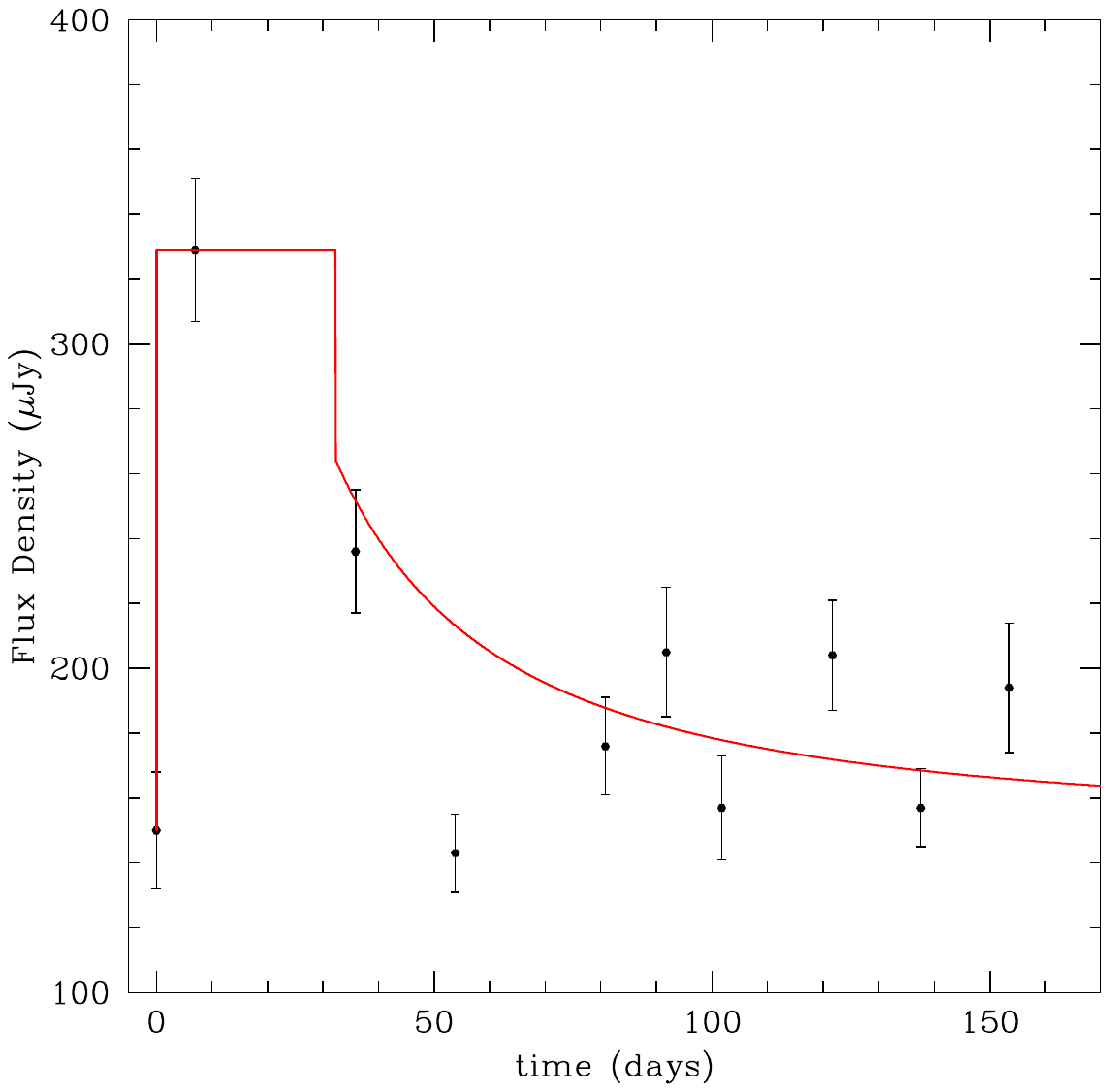}
\vskip-4.0cm
\caption{Radio continuum observations of V2187 Cyg at a frequency
of 4.9 GHz (black dots); shock model of the radio pulse (red solid line).
A discussion of the figure is given in the text.}
\label{fig:flujo_ala.eps}
\end{figure}

Choosing a velocity variation factor $a=1.5$ and duration of the constant velocity phase $t_c = 32.3$ days, 
from a least squares fit of the observed radio pulse,
one finds that the mass-loss rate variation factor is $b=4.65$, and the ejection time is $\Delta \tau \sim 3.90$  days
\footnote{Note that we have ignored one point (the fourth epoch)
in the least square fit of the radio pulse.}.
The emission of this model is presented in Figure~\ref{fig:flujo_ala.eps} with a red solid line, together with the observations.
%Initially, when the WS moves with a constant velocity, the flux is constant, and, when the WS decelerates, the flux density decreases with time.

We now assume that a constant fraction $\epsilon$  of the bolometric luminosity, is radiated at $4.9$ GHz,
$S_{WS}\,$(4.9\,$\mbox{GHz}$)= $\epsilon \,L_{WS}/ (4\pi D^2)$, where the distance is $D=1.78$ kpc. 
In Table \ref{tab:param} we present the parameters for main sequence early B-type stars with solar abundances.
The stellar temperature $T_\star$ (column 2), luminosity $L_\star$ (column 3), radius $R_\star$ (column 4) and rate of ionizing photons
$\dot N_i$ (column 4)  are taken from \cite{1973AJ.....78..929P} . 
The ionizing photon rate required to ionize the wind $\dot N_{\rm wind}$ (column 6) is derived from the formulation of  \citetalias{1983RMxAA...8..163R}
(note that in their equation 3, the exponent of the last term should be $[R/10~R_{\odot}]^{-1}$).
Their formulation assumes purely ionized hydrogen gas at an electron temperature of $10^4$ K.
Since this rate is orders of magnitude smaller than the actual
stellar rate, these stars easily ionize their associated winds.
The stellar mass $M_\star$ (column 7) is derived from the mass-luminosity relation for OB stars
given by  \cite{2007AstL...33..251V}, while the mass-loss rate $\dot M$ (column 8) is that given by the 
relation of \cite{1990A&A...231..134N}. 
Finally, the terminal wind velocity $v_\infty$ (column 9) is obtained from a modified version of the prescription by \cite{2000ARA&A..38..613K}.
This prescription relates the terminal velocity to the escape velocity with an expression of the form
$v_\infty = C(T_\star) v_{esc}$, where $v_{esc} = \sqrt{ G M_\star / R_\star}$,  and has a discontinuity at $T_\star$ = 21,000 K. We have substituted this discontinuity with
a smooth step function of the form 
$$C(T_\star) = 2.03  +  0.63~\tanh \biggl[ {{T_\star - 21,000 K} \over {a \times21,000 K}} \biggr],$$
{ where $a=0.2$ was chosen to obtain a gradual transition.}

For a B3V star like V2187 Cyg,  Table \ref{tab:param} gives a wind
mass-loss rate $ \dot{M}$= 4.3 $\times 10^{-10}\, M_{\odot}\, yr^{-1}$ and a velocity $v_w= 1090 ~\, {\rm km~\, s^{-1}}$. 
In this case,  the
maximum bolometric flux of the model is $S_{WS}=2.17 \times 10^{9}$ Jy (see equation \ref{eq:lumi}), 
while the observed  peak flux of the radio pulse at 4.96 GHz, over a baseline of $\sim 150~ \mu$Jy,
is $S(4.9 \, {\rm GHz}) \sim 179~
 \mu$Jy. Comparing both fluxes one obtains a fraction 
$\epsilon= 8.24 \times 10^{-14} \left({4 \pi \over \Omega}\right)$ for the bolometric luminosity of the WS that is radiated at 4.9 GHz. 
If one takes, instead, the maximum mass-loss rate of an 
ionized wind for a B3V star (see Table  \ref{tab:param2}), $ \dot M_{max}= 1.2 \times 10^{-7}\, M_{\odot}\, yr^{-1}$, the bolometric flux increases to 
$S_{WS}=6.06 \times 10^{11}$ Jy, and the fraction of energy radiated at 4.9 GHz decreases to 
$\epsilon= 2.95 \times 10^{-16} \left({4 \pi \over \Omega}\right)$.
Therefore, in both
mass loss rates discussed here there is enough energy in internal shocks of the stellar wind to produce the observed radio pulse at 4.9 GHz in V2187 Cyg. Nevertheless, since this simple model does not solve for the shock microphysics, the  emission of the WS could be thermal or non thermal.

%\noindent{\bf \color{red} Table 2 here}

\begin{deluxetable*}{cccccc}
\tablenum{3}
\tablecaption{Derived Parameters for Main Sequence Early-Type B Stars}
\tablewidth{900pt}
\tabletypesize{\scriptsize}
\tablehead{
\colhead{}    & \colhead{S$_{\rm thin}$}    & \colhead{S$_{\rm wind}$}    & \colhead{$\dot {\rm M}_{\rm max} $}    &  \colhead{} & \colhead{$S_{\rm max}$}    \\
\colhead{Sp}    & \colhead{(mJy)}    & \colhead{(mJy)} &\colhead{(M$_\odot$ yr$^{-1}$)} & \colhead{$\dot {\rm M}_{\rm max} $/$\dot {\rm M}$}    &   \colhead{(mJy)}   }
\decimalcolnumbers
       \startdata
      B0V & 4.4$\times$10$^{3}$ &
      9.0$\times$10$^{-4}$ & 2.5$\times$10$^{-5}$  & 480 &  3.4$\times$10$^{+0}$ \\ 
     B0.5V & 3.5$\times$10$^{2}$ &
        2.2$\times$10$^{-4}$ & 5.8$\times$10$^{-6}$   & 390  & 6.1$\times$10$^{-1}$ \\
       B1V & 3.5$\times$10$^{1}$ &  
        5.1$\times$10$^{-5}$   & 1.4$\times$10$^{-6}$ & 290  & 1.2$\times$10$^{-1}$ \\
       B2V & 7.0$\times$10$^{0}$ &    
       4.7$\times$10$^{-5}$ &   4.7$\times$10$^{-7}$  & 210 & 3.8$\times$10$^{-2}$ \\
        B3V & 8.8$\times$10$^{-1}$ &  5.8$\times$10$^{-6}$   &  1.2$\times$10$^{-7}$  & 280  & 8.1$\times$10$^{-3}$ \\ 
        \enddata  
\tablecomments{Flux densities are at 4.96 GHz for a star located at 1 kpc. Sp = Spectral Type, S$_{\rm thin}$ = Flux density for optically-thin case, S$_{\rm wind}$ = Flux density for ionized wind, $\dot {\rm M}_{\rm max} $ = Maximum mass loss rate that can be fully ionized by the star,
$\dot {\rm M}_{\rm max} $/$\dot {\rm M}$ = Ratio of the maximum mass loss rate to the 
nominal mass loss rate given in Table \ref{tab:param}, $S_{\rm max}$ = Maximum flux density for an ionized stellar wind.}
\label{tab:param2}
\end{deluxetable*}

%\noindent{\bf \color{red} Table 3 here}

\begin{deluxetable*}{ccccc}
\tablenum{4}
\tablecaption{Upper limits for the radio emission of $\beta$ Cep}
\tablewidth{900pt}
\tabletypesize{\scriptsize}
\tablehead{
\colhead{}  & \colhead{Gain}  & \colhead{Frequency}  & \colhead{Bandwidth}  & \colhead{Upper limit} \\
\colhead{Epoch} & \colhead{Calibrator} & \colhead{(GHz)} & \colhead{(GHz)} & \colhead{($\mu$Jy)} }
\decimalcolnumbers
       \startdata
2014 Apr 24 &  J2005+7752 & 3.0 & 2.0 &  $\leq$29 \\
2014 Apr 22  & J2005+7752 & 10.0 & 4.0 & $\leq$18  \\
2014 Apr 22 & J2009+7229  & 33.0 & 8.0  & $\leq$31 \\
\enddata  
\label{tab:param4}
\end{deluxetable*}

\subsection{Origin of the quiescent emission: an ionized wind origin?}

The first point of Figure \ref{fig:flujo_ala.eps} shows a baseline level of emission at 4.9 GHz of $F \sim 150 \, \mu$Jy. This emission has a spectral 
index $\alpha \sim 0.6$, characteristic of  a stellar wind. Here we examine  the possibility that the quiescent emission comes from the 
ionized stellar wind, discussed in the previous section.

In Table \ref{tab:param2}  we list the derived parameters for early-type MS B stars using the formulation of \citetalias{1983RMxAA...8..163R} . 
We normalize the calculations assuming that the stars are located at 1 kpc.
The flux density for an optically-thin free-free plasma $S_{\rm thin}$ (column 2), derived with equation (2) of
\citetalias{1983RMxAA...8..163R}, is orders of magnitude larger than the flux density for an
ionized wind $S_{\rm wind}$ (derived with equation (1) of \citetalias{1983RMxAA...8..163R}) with the parameters given in Table \ref{tab:param}. The maximum flux density $S_{\rm max}$  corresponds to an ionized wind where the mass loss rate is large enough to
fully absorb all stellar ionizing photons. This maximum mass loss rate $\dot M_{\rm max}$ (in column 4),
is above two orders of magnitude larger that the ``nominal" mass loss rate $\dot M$ given in Table  \ref{tab:param}. In other words,
while $\dot M$ is the mass loss rate usually attributed to these spectral types (the ``nominal" mass loss rate), 
$\dot M_{\rm max}$ is the hypothetical mass
loss required to exactly absorb all the stellar ionizing photons. Are the large mass loss rates $\dot M_{\rm max}$
possible? The analysis of observational mass loss rates of B stars
by \cite{1985ApJ...294..567V}  and \cite{2019AJ....158...73K}  imply large dispersions of several 
orders of magnitude.
We will then leave as an open possibility the existence of large mass loss rates, comparable to those listed in column 4 of this Table.

\begin{figure*}[!t]
\vskip-2.8cm
%\hskip-0.4cm
%\includegraphics[width=0.92\linewidth]{Bstars.pdf}
\includegraphics[width=\linewidth]{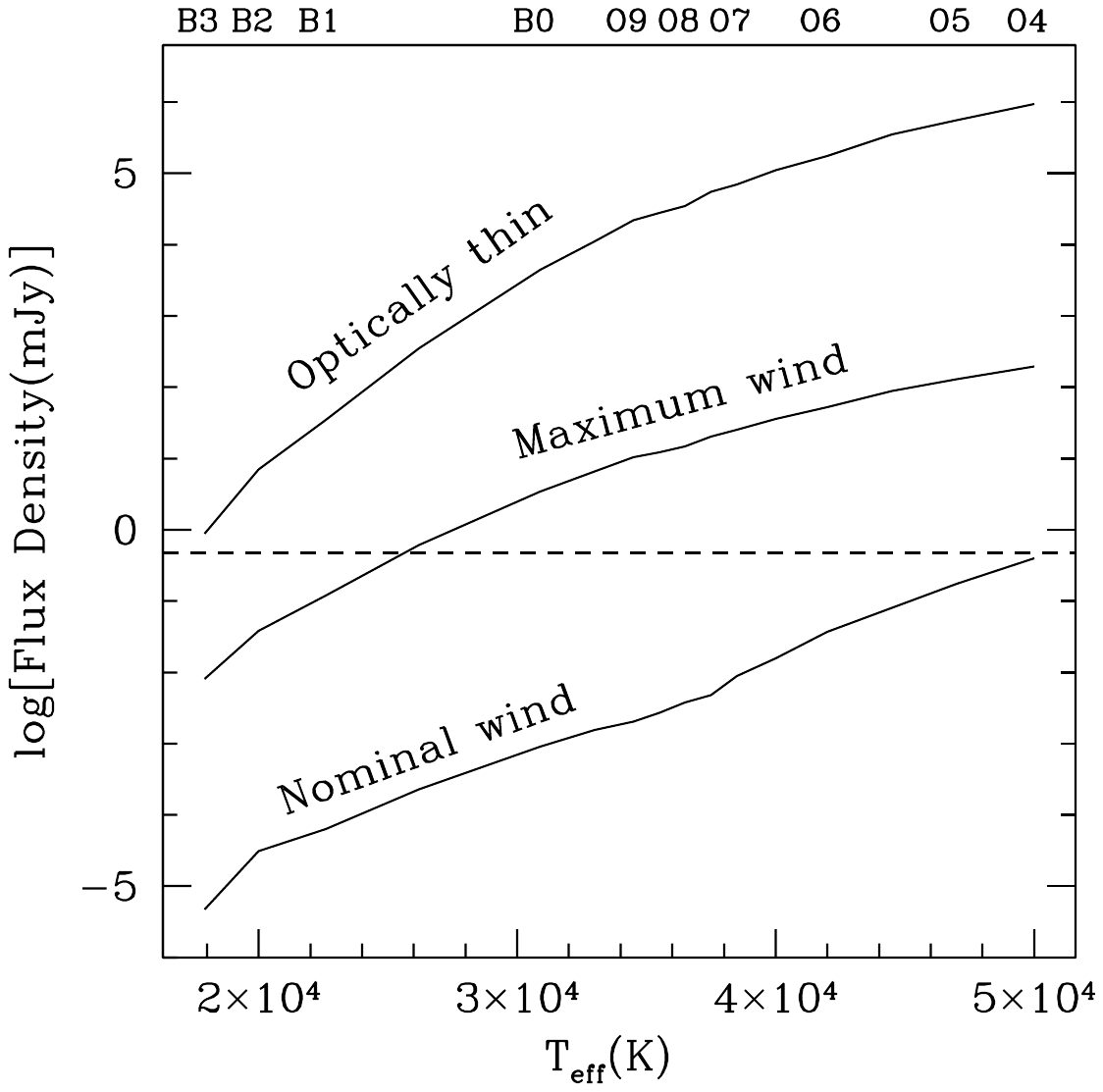}
\vskip-4.8cm
\caption{%\scriptsize 
Flux densities of the three models discussed as a function of the effective
temperature of the star (or alternatively, the spectral type, as indicated in the top horizontal axis). 
The dashed horizontal line represents the flux density of V2187 Cyg
normalized to a distance of 1 kpc.
}
\label{fig:fluxes}
\end{figure*}

In Figure \ref{fig:fluxes} we plot the three different flux densities $S_{\rm wind}$, $S_{\rm max}$, and $S_{\rm thin}$, 
as a function of the effective temperature of the star (or alternatively,
the spectral type). We have extended the temperature range to more luminous stars (up to O4) using the parameters
in Table 3 of \cite{2013ApJ...763..139D}.
The horizontal dashed line marks the 4.96 GHz flux density of V2187 Cyg  (0.15 mJy) normalized to a distance of 1 kpc (0.48 mJy). From this Figure we see that this flux density can be explained with optically-thin free-free emission
from any spectral type. But this emission will have a spectral index of -0.1, in disagreement with the value of 0.6 determined by us.
The maximum wind model could explain the quiescent emission at 4.9 GHz  but for a spectral type of B0.5 or earlier. The nominal wind model fails by orders of magnitude to account for the observed flux density. 

We conclude that we cannot explain the quiescent radio emission of V2187 Cyg in terms of thermal emission from the nominal stellar wind.

\subsection{External ionization?}

As discussed above, there is a maximum mass-loss rate that can be ionized by the ionizing flux of the OB stars. 
If the mass-loss rate of the star, $\dot M$, is higher that $\dot M_{\rm max}$ in Table \ref{tab:param2}, the ionization front would be trapped close to the stellar
surface, and the wind would be largely
neutral.  Here we consider the possibility that V2187 Cyg has a high mass-loss rate such that the wind 
is neutral, and is externally ionized by a nearby O star.

There is a nearby O4.5III star (Cyg OB2 8C) projected at 51.88 arcsec (corresponding to 0.45 pc at a distance of 1.78 kpc)
from V2187 Cyg. According to Panagia (1973) this star would have an ionizing rate of $6.6 \times 10^{49} \, {\rm s}^{-1}$, 
producing an ionizing flux $F_{O4} = 2.76 \times 10^{12} {\rm \, cm^2 \, s^{-1}}$ impinging on the neutral wind.  
The ionization front is formed at a distance $r_0$ from the star with the neutral wind, along the symmetry axis that connects both stars,  given by the condition that the flux of ionizing photons equals the flux of neutral particles from the stellar wind,
\begin{equation}
F_{O4} \sim n_w v_w \sim {\dot M_w \over 4 \pi r_{0}^2 \mu m_H} ,
\end{equation}
Cant\'o et al. (2024; in preparation).
In addition, the optical depth of the ionized gas along the symmetry axis  is given by
\begin{equation}
\tau_{\nu} = \chi A^2  \int_{r_0}^\infty {dr \over r^4}  ={ \chi _\nu A^2  \over 3 r_0^3}, 
\end{equation}
where the free-free opacity is $\chi_\nu =  8.436 \times 10^{-28} \left( { \nu \over 10 GHz}\right)^{-2.1} \left( { T \over 10^4 K } \right)^{-1.35} {\rm cm^5}$, and the density coefficient is $A =  {\dot M_w / ( 4 \pi v_w \mu m_H)}$.
For an externally ionized wind of a B3V star with a speed of $v_w = 1090 \, {\rm km s^{-1}} $ (Table \ref{tab:param}),  and a mass-loss rate of $\dot M_w = 1.25 \times 10^{-5} M_\odot {\rm yr^{-1}}$, the optical depth is $\tau_{4.9 GHz} \sim 2.6 \times 10^{-3} << 1$. For these parameters, the model of  Cant\'o et al. (2024), gives a flux of the ionized gas of the order of the observed flux, $F_{4.9GHz}  \sim 150 \, \mu$Jy.
Nevertheless, as shown above, the emission of this externally ionized wind would be optically thin, i.e., with a flat spectrum, in contrast with the measured spectral index of  V2187 Cyg, $\sim 0.6$. The latter spectral index can appear in the case of free-free emission from a  partially optically thick ionized stellar wind, or optically-thick synchrotron or gyrosynchrotron emission. 

Thus, for the case discussed above, the quiescent radio emission of V2187 Cyg cannot be due to a neutral stellar wind that is externally ionized by a nearby O star. It should be pointed out, however,
that the optical depth (and, correspondingly, the spectral index)
can increase for much larger values of the ionizing flux or by increasing the wind density constant $A$. Nevertheless, in this case, recombinations in the ionized wind need to be taken into account.

\subsection{A magnetic mechanism?}
We have discussed how the observed 4.9 GHz continuum baseline of 0.15 mJy of V2187 Cyg is too high to be due to the free-free
emission of the stellar wind of a B star or to the external ionization of a massive neutral wind by a nearby O star. 
Another possibility is that the observed radio emission has a non thermal origin.  Relatively
strong radio emission is observed in early-type magnetic stars (e.g., \cite{1987ApJ...322..902D}, \cite{1992ApJ...393..341L}, \cite{1994A&A...283..908L}).
{The specific radio luminosity of V2187  Cyg at 4.9 GHz  is $L_{4.9 {\rm GHz}} = 5.69 \times 10^{17}  {\rm erg \, s^{-1} \, Hz^{-1}}$, which is within the range of emission 
of MS magnetic B stars (see, e.g.,  Table 1 of \cite{2021MNRAS.507.1979L}). 
These early-type magnetic stars have magnetic fields of the order of several  kGauss.} The  large scale magnetic field
affects the stellar wind: the wind flow is magnetically confined in the equatorial regions and flows freely in the poles, as described by the magnetically confined wind shock model (MCWS) of \cite{1997A&A...323..121B}. 
This model successfully explains the spectra of early-type magnetic stars from the radio to the X rays. Non-thermal radio emission is  produced by relativistic electrons  accelerated in equatorial current sheets 
that interact with the stellar magnetosphere and produce  partially circularly polarized gyro-synchrotron radiation.
In the framework of the MCWS model, radio light curves and spectra have been calculated, for example, by 
\cite{2004A&A...418..593T} and  \cite{2006A&A...458..831L}. %, and \cite{2021MNRAS.507.1979L} . 
Although, in general, the modeled radio spectra between 1 and $\sim$ 30 GHz is approximately flat,
 the spectral index  $\alpha  \sim 0.6$  in the range 1-5 GHz  is also possible both in observed and modeled spectra of MS magnetic B stars (see, e.g, Figure 2 of \cite{2021MNRAS.507.1979L}). 
 
{Thus, the quiescent radio emission of V2187 Cyg could be non-thermal.  }To test the possibility that V2187 Cyg is a magnetic star, one needs to look for circular polarization in the radio
as expected from gyro-synchrotron radiation and also try to measure the stellar magnetic field, that
is expected to be in the range of several kGauss.

%\section{Time Behavior of the Radio Emission}

\section{Upper limits to the radio flux density of $\beta$ Cep}
\label{sec:origin}

As discussed by Tapia et al. (2014), V2187 Cyg is the only $\beta$ Cep star with detected radio continuum emission. These authors give a list of 15 additional $\beta$ Cep stars with sensitive VLA observations and 3-$\sigma$ upper limits in the range of 0.1 to 0.6 mJy. In particular, the prototype of the class, $\beta$ Cep, had an upper limit of 0.15 mJy at a frequency of 8.3 GHz for the epoch 2002.42.

We searched for more recent observations of $\beta$ Cep in the VLA archive, finding sensitive observations as part of the project 14A-139. These observations were made in the A configuration and J0137+331 (3C48) was always the amplitude calibrator. In this project $\beta$ Cep was observed at three frequencies, obtaining the 3-$\sigma$ upper limits given in Table 4.
We also searched unsuccessfully for pulsed emission binning the data in time intervals from 10 to 120 seconds. $\beta$ Cep hosts a sinusoidally varying magnetic field that reaches amplitudes of $\sim$100 G (Henrichs et al. 2013), but this field seems to be insufficient to
produce detectable radio emission. The stars with radio emission discussed by  \cite{2021MNRAS.507.1979L} have characteristically
magnetic fields of the order of several kG and fields this large may be required to produce detectable radio emission.

Interpolating the  $\beta$ Cep spectrum to 4.96 GHz, we find that this upper limit
is about 10 times smaller than the 5.0 GHz flux density discussed here for V2187 Cyg. Furthermore, 
$\beta$ Cep is at a distance on only 210 pc, while V2187 Cyg is at 1.78 kpc. In conclusion, V2187 Cyg is at least 700 times more radio luminous than $\beta$ Cep. This large ratio could be understood as follows. The emissivity of magnetic processes goes
as the magnetic field squared, $B^2$. If V2187 Cyg has a magnetic field of a few kG, its radio emission will be about $10^3$ times 
larger than that of $\beta$ Cep. To test this hypothesis a determination of the magnetic field of V2187 Cyg
is required.

%\noindent{\bf \color{red} Table 4 here}

\section{Conclusions}

1) We analyzed archive VLA observations of the $\beta$ Cep star V2187 Cyg. The observations were made in 10 epochs at 
1.39 and 4.96 GHz. The spectral index obtained from images made concatenating all epochs is $\alpha = 0.6\pm0.2$ ($S_{\nu} \propto \nu^\alpha$), consistent with an ionized wind 
 or a partially optically-thick synchrotron or gyrosynchrotron source.
 
 2) A temporal pulse is observed in the 4.96 GHz data that can be explained in terms of shocks propagating
in the stellar wind of V2187 Cyg. 

3) The quiescent radio emission of V2187 Cyg at 4.96 GHz (with a flux density of $\simeq 150~\mu Jy$), cannot be explained in terms 
of an internally or externally photoionized wind.

4) We conclude that the radio emission of V2187 Cyg most likely has a magnetic origin.  This can be tested with
a sensitive search for circular polarization in the radio
as expected from gyro-synchrotron radiation and also by trying to measure the stellar magnetic field, that
is expected to be in the range of several kGauss. 

\begin{acknowledgments}

%\section{Acknowledgements}

We thank the anonymous referee for comments that improved the clarity of our paper.
L.F.R. acknowledges the financial support of PAPIIT - UNAM 
IN108324 and CONAHCyT 238631.
S. L., J. C., and R. F. G., acknowledge support from  PAPIIT - UNAM IN102724,  IG100422, and IN103023, respectively.
M.T. acknowledges financial support from PAPIIT-UNAM IN110422.

\end{acknowledgments}

\end{document}